\documentclass[twocolumn,times]{aastex62}

\usepackage{makecell}

\begin{document}

\title{\Large\bf Rapid Flaring in the Galactic-plane Gamma-ray Transient Fermi J0035+6131}

\shorttitle{Rapid Flaring in Fermi J0035+6131}

\author[0000-0003-2085-5586]{Dirk Pandel}
\affiliation{Grand Valley State University,
118 Padnos Hall of Science, 1 Campus Drive,
Allendale, MI 49401, USA}

\author[0000-0002-3638-0637]{Philip Kaaret}
\affiliation{The University of Iowa,
Department of Physics and Astronomy,
Iowa City, IA 52242, USA}

\shortauthors{Pandel \& Kaaret}

\journalinfo{Submitted to ApJ, Published version at \textup{\url{https://doi.org/10.3847/1538-4357/aacbc0}}}

\tabletypesize{\footnotesize}


\begin{abstract}

We investigate the gamma-ray and \mbox{X-ray} emission from the transient gamma-ray
source Fermi J0035+6131, which was discovered with the \textsl{Fermi} Large Area
Telescope (LAT) near the Galactic plane at $b=1\fdg3$,
and we discuss potential multi-wavelength counterparts of the gamma-ray source.
Our analysis of over 9~years of \textsl{Fermi} LAT data revealed two flaring events
lasting \mbox{10--30 hr} during which the gamma-ray flux increased by a factor
of $>$300 compared to the long-term average.
We also analyzed \mbox{X-ray} data obtained with \textsl{XMM-Newton} and \textsl{Swift}
and identified several sources with a hard \mbox{X-ray} spectrum
inside the \textsl{Fermi} LAT confidence region.
The two brightest \mbox{X-ray} sources have known counterparts at other wavelengths
and are associated with the compact radio source VCS4 J0035+6130 and
the \mbox{B1 IV:nn} star \mbox{HD 3191}, respectively.
VCS4 J0035+6130, which is also detected in the near infrared,
is likely an active galaxy serendipitously located behind the Galactic disk
and is the most compelling candidate for the counterpart of the gamma-ray source.
\mbox{HD 3191} appears to be part of an \mbox{X-ray} binary with a compact companion
and is unlikely to be associated with Fermi J0035+6131.

\end{abstract}



\section{Introduction}

\setcounter{footnote}{2}

The Large Area Telescope (LAT) on board the \textsl{Fermi} Gamma-ray Space Telescope
\citep{2009ApJ...697.1071A} has detected numerous, previously unknown transient sources
of GeV gamma rays.
While the majority of these transient gamma-ray sources are likely extragalactic, 
some have been shown to be located within our Galaxy.
One such Galactic gamma-ray transient is Fermi J2102+4542 \citep{atel2487}, which was
identified as a classical nova outburst in the symbiotic star \mbox{V407 Cyg}
\citep{2010Sci...329..817A}.
This discovery established novae as a new class of gamma-ray sources, and several other novae
have subsequently been detected with the \textsl{Fermi} LAT. 
Variable GeV gamma-ray emission has also been observed from several Galactic \mbox{X-ray}
binaries such as the microquasar Cygnus \mbox{X-3} \citep{2009Sci...326.1512F} and the
High-mass \mbox{X-ray} Binaries \mbox{LS I +61 303} and \mbox{LS 5039}
\citep{2009ApJS..183...46A}.
An investigation of \textsl{Fermi} LAT transients detected near the Galactic plane may reveal
previously unknown GeV gamma-ray sources in our Galaxy.

On 2016 January 14, the \textsl{Fermi} LAT observed strong gamma-ray emission from the
previously unknown source Fermi J0035+6131 located only $1\fdg3$ from the Galactic plane.
The source was detected with a daily-averaged flux of
$(5.7\pm1.5)\times10^{-7}\ \mathrm{photons\ cm^{-2}\ s^{-1}}$ above \mbox{100 MeV}
at coordinates $\mathrm{R.A.}=8\fdg91$ and $\mathrm{decl.}=61\fdg52$ (J2000.0) with a statistical
95\% containment radius of $0\fdg08$ \citep{atel8554}.
The best-fit location of the gamma-ray source is $0\fdg03$ from the unidentified radio source
VCS4 J0035+6130 (87GB 003232.7+611352) which was suggested as a potential counterpart.
Follow-up observations of this source in the radio band \mbox{6--23 days} after the gamma-ray
detection indicate that, compared to earlier observations, the radio flux was higher
by a factor of \mbox{1.5--2} at frequencies above \mbox{4.8 GHz} \citep{atel8718}.
A near-infrared (NIR) counterpart of the radio source with magnitudes
$J=16.81$, $H=15.64$, and $Ks=14.86$ was discovered in an observation
performed 14~days after the gamma-ray detection \citep{atel8706}.

The region near Fermi J0035+6131 was also observed in \mbox{X-rays} with \textsl{XMM-Newton}
10~days after the \textsl{Fermi} LAT detection.
In a preliminary analysis of the \textsl{XMM-Newton} data, we detected two previously unknown
\mbox{X-ray} sources inside the \textsl{Fermi} LAT confidence region, the radio source
VCS4 J0035+6130 mentioned above and the \mbox{B1 IV:nn} star \mbox{HD 3191}
\citep{atel8783,2017AIPC.1792d0037P}.
\citet{atel8789} obtained optical spectra of \mbox{HD 3191} and derived its radial velocity,
which was found to differ significantly from that measured several decades earlier.
The hard \mbox{X-ray} spectrum and the apparent change in radial velocity suggest that
\mbox{HD 3191} may be an \mbox{X-ray} binary.
Both \mbox{X-ray} sources are potential candidates for the multi-wavelength counterpart
of Fermi J0035+6131.

In this paper, we present a detailed analysis of gamma-ray and \mbox{X-ray} data
of the region near Fermi J0035+6131.
We analyze over 9~years of \textsl{Fermi} LAT Pass~8 data to search
for additional gamma-ray flaring activity, obtain an improved location of the transient,
and constrain its spectral properties.
Furthermore, we analyze \mbox{X-ray} data obtained with \textsl{XMM-Newton} and \textsl{Swift} 
to identify potential counterparts of Fermi J0035+6131 and constrain their \mbox{X-ray} spectra
and variability.
We then investigate the multi-wavelength properties of these counterparts
and discuss the implications for the identification of the gamma-ray transient.


\section{\textsl{Fermi} LAT Data Analysis}

We analyzed \textsl{Fermi} LAT Pass~8 data \citep{2013arXiv1303.3514A}
covering 9.5~years from 2008 August~4 to 2018 March~2.
Events were selected in the energy range \mbox{0.1--100 GeV} from a $15\degr$ radius circular
region of interest centered on $\mathrm{R.A.}=8\fdg91$ and $\mathrm{decl.}=61\fdg52$,
the best-fit location of Fermi J0035+6131 originally reported by \citet{atel8554}.
Included in our analysis were front and back events classified as \texttt{SOURCE}
(\texttt{evclass=128}) with zenith angles $<$90$\degr$,
and the Instrument Response Functions \texttt{P8R2\_SOURCE\_V6} were used.
The spatial, spectral, and timing analysis of the LAT data was performed using a standard
unbinned maximum likelihood method with the Fermi Science
Tools.\footnote{\url{https://fermi.gsfc.nasa.gov/ssc/data/analysis/}}
The significance of gamma-ray sources was evaluated using the Test Statistic
$\mathrm{TS}=2\ln(\mathcal{L}_1/\mathcal{L}_0)$ \citep{1996ApJ...461..396M},
where $\mathcal{L}_1$ and $\mathcal{L}_0$ are the maximum likelihood values
with and without a given source included in the model.
$\sqrt{\mathrm{TS}}$ is approximately equal to the significance in terms of the standard deviation $\sigma$.
In a preliminary analysis of two days of LAT data of the transient event (MJD 57401--57403)
we detected four of the sources listed in the 3FGL catalog \citep{2015ApJS..218...23A}
inside the $15\degr$ radius region of interest with a Test Statistic \mbox{$\mathrm{TS}>10$}:
3FGL J0007.0+7302, 3FGL J0223.6+6204, 3FGL J0240.5+6113, and 3FGL J2229.0+6114.
We included these sources in our model for the likelihood analysis,
but because of their comparatively low number of counts and their large distances
from Fermi J0035+6131 of $>$10$\degr$, the spectral parameters were fixed at the catalog values.
Our model of the region of interest also included Galactic diffuse emission (\texttt{gll\_iem\_v06.fits})
and extragalactic isotropic emission (\texttt{iso\_P8R2\_SOURCE\_V6\_v06.txt})
with their normalizations as free parameters.

\begin{figure}
\includegraphics[width=0.97\linewidth]{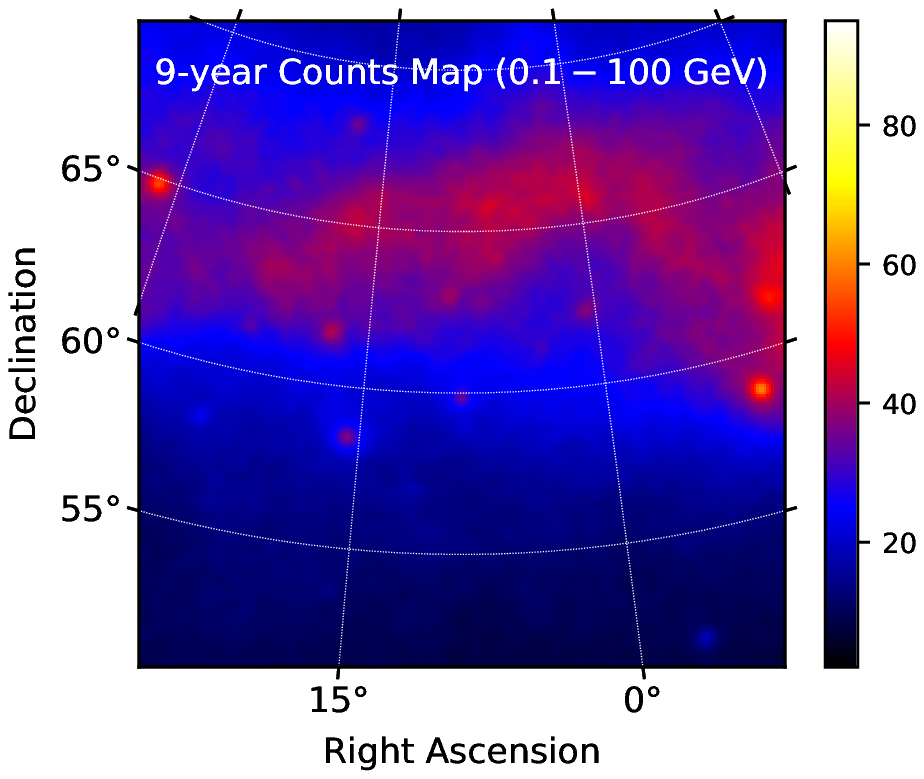}
\includegraphics[width=0.97\linewidth]{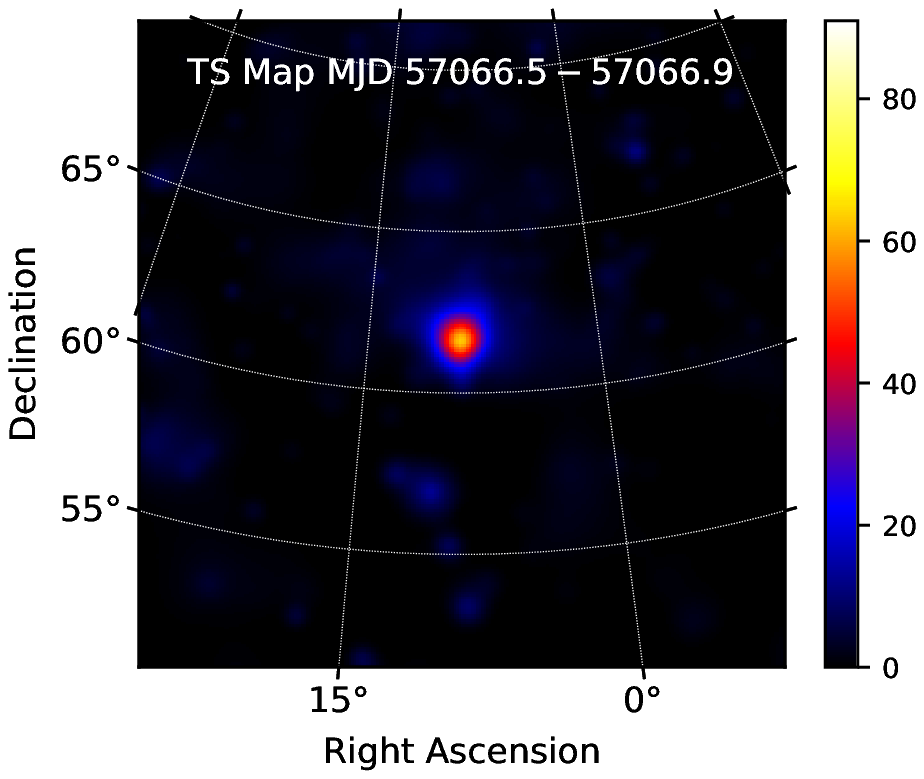}
\includegraphics[width=0.97\linewidth]{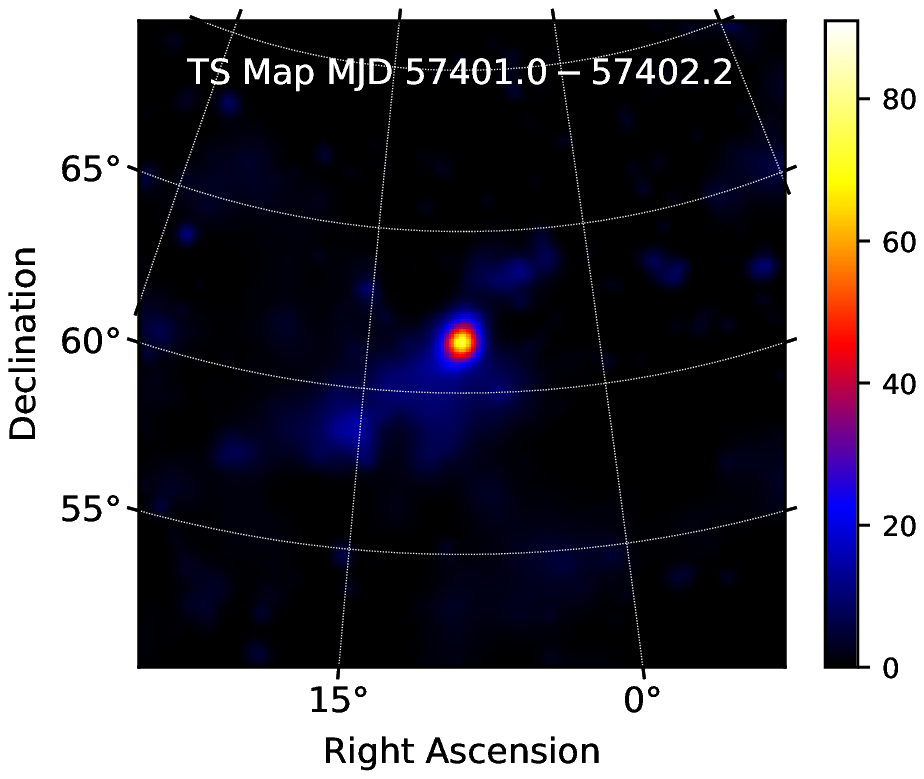}
\caption{
Smoothed \mbox{9-year} gamma-ray counts map (bin size $0\fdg05$)
of a $20\degr\times20\degr$ region centered on Fermi J0035+6131
and TS maps at the times of the two flares.
\label{gamma-map}}
\end{figure}

\begin{figure*}
\hfill
\includegraphics[width=0.45\linewidth]{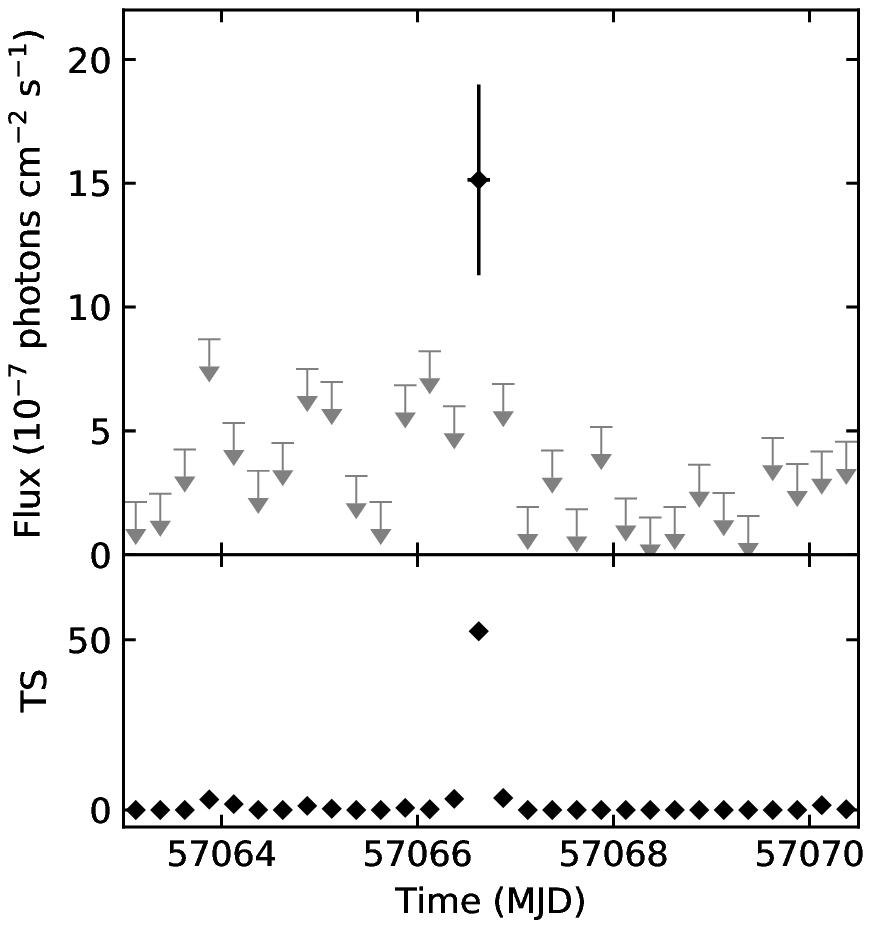}
\hfill
\includegraphics[width=0.45\linewidth]{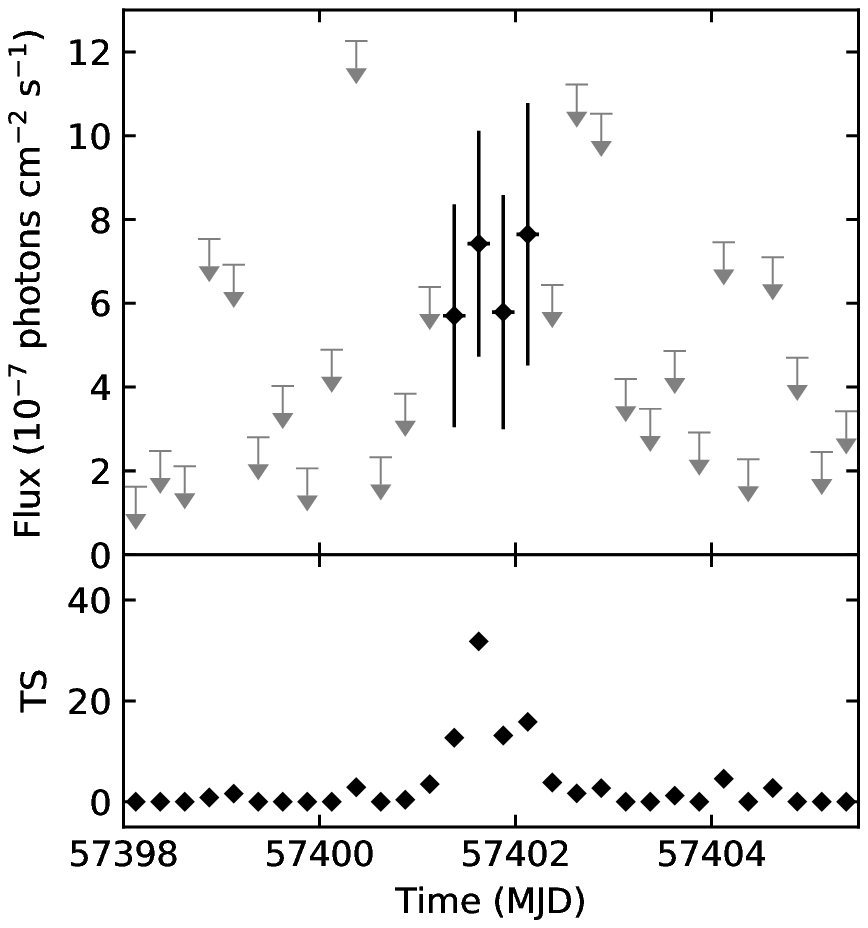}
\hfill
\caption{
\mbox{6-hr} gamma-ray light curves showing the flux (\mbox{0.1--100 GeV}) and the Test
Statistic TS of Fermi J0035+6131 near the times of the two flares.
Error bars are shown at 68\% confidence and upper limits at 95\% confidence.
\label{gamma-lc}}
\end{figure*}

A gamma-ray counts map of 9.5 years of LAT data of the region around Fermi J0035+6131
is shown in Figure~\ref{gamma-map}.
No gamma-ray source is detected at the location of the transient, which places a
95\% upper limit of  $1.5\times10^{-9}\ \mathrm{photons\ cm^{-2}\ s^{-1}}$
on the average gamma-ray flux between 0.1 and \mbox{100 GeV} (assuming a power-law spectrum
with a photon index of $-2.0$).
To search for previously undetected gamma-ray flaring activity in Fermi J0035+6131,
we divided 9.5~years of LAT data between MJD 54683 and MJD 58180
into \mbox{2-day} time intervals in \mbox{1-day} steps and fitted our model of the region of interest 
to the data of each time interval.
Fermi J0035+6131 was modeled as a point source at its originally reported location
having a power-law spectrum with a fixed photon index of $-2.0$ and variable normalization.
We searched the light curve for time intervals during which the gamma-ray source is detected
with a TS value $>$25, which corresponds to a $5\sigma$ detection threshold for a single
\mbox{2-day} interval or a 0.2\% probability of at least one false detection in the entire
light curve.
We found two flaring events above this threshold with TS values of 35 and 82, respectively,
with the lower value corresponding to a false-detection probability of $1\times10^{-5}$.
The first flare occurred near MJD 57067 and had not previously been reported.
The second flare, which was detected near MJD 57402, is the one reported by \citet{atel8554}.
Figure~\ref{gamma-lc} shows \mbox{6-hr} light curves of the gamma-ray flux
and the Test Statistic TS near the times of the two flares.

To better constrain the start and end times of the two flares,
we selected gamma-ray events from a $1\degr$ region around Fermi J0035+6131
and used a Bayesian Blocks method \citep{2013ApJ...764..167S}
with the unbinned event data.
We found that the first flare lasted \mbox{$\sim$10 hr} from MJD 57066.5 to 57066.9
and the second flare \mbox{$\sim$30 hr} from MJD 57401.0 to 57402.2
with the first flare being three times as intense as the second.
(Note that the start and end times could not be constrained to better than $\sim$0.1~days
because the object was only observable during $\sim$1/4 of the \mbox{3.2-hr}
\textsl{Fermi} LAT sky survey cycle.)
We then selected LAT data from the entire region of interest for the above time intervals
and fit our model to the data using an unbinned maximum likelihood method.
Figure~\ref{gamma-map} shows TS maps for the two time intervals obtained
using \texttt{gttsmap}.
The TS maps, which represent the significance of a putative point source at a given location,
clearly show the detection of Fermi J0035+6131 with a high significance.
The best-fit location of the gamma-ray source, i.e.\ the location with the maximum TS value,
was determined using \texttt{gtfindsrc}.
Results are shown in Table~\ref{gamma-table}.
To obtain stronger constraints on the source location, we also performed an analysis
of the combined data from the two time intervals.
Our best estimate for the location of Fermi J0035+6131 is $\mathrm{R.A.}=8\fdg975$ and
$\mathrm{decl.}=61\fdg548$ with a location error of $0\fdg056$ (68\% confidence, statistical only).

The spectral parameters of the two flares were obtained using an unbinned maximum likelihood fit
with a simple power-law model with the normalization and photon index as free parameters.
The average gamma-ray flux and the photon index obtained from the fit
are shown in Table~\ref{gamma-table}.
The spectra are well described by a power-law model with a photon index near $-2.0$.
The number of detected gamma rays was too low to allow discriminating
between different spectral models such as broken power-law or log-parabolic models.
Given the upper limit of $1.5\times10^{-9}\ \mathrm{photons\ cm^{-2}\ s^{-1}}$
on the \mbox{9.5-year} averaged gamma-ray flux,
the two flares represent an increase in gamma-ray brightness by factors of at least
1000 and 300, respectively.

\begin{deluxetable*}{lccccccc}
\setlength{\tabcolsep}{1em}
\tablecaption{
Best-fit location, detection significance, source counts predicted by the model,
and spectral parameters of Fermi J0035+6131 from our analysis of the individual
and combined \textsl{Fermi} LAT data of the two gamma-ray flares.
\label{gamma-table}}
\tablehead{
Time Interval  &  \multicolumn{2}{c}{Best-fit Location}  &  \colhead{Location}      &  \colhead{TS Value}       &  \colhead{Predicted}  &  \colhead{Flux (0.1--100 GeV)}              &  \colhead{Power Law}  \\[-1ex]
\cline{2-3} \\[-4ex]
(MJD)          &  \colhead{R.A.}  &  \colhead{Decl.}     &  \colhead{Error (68\%)}  &  \colhead{(Significance)}  &  \colhead{Counts}     &  \colhead{($\mathrm{photons\ cm^{-2}\ s^{-1}}$)}  &  \colhead{Photon Index}
}
\startdata
57066.5--57066.9  &  $8\fdg993$  &  $61\fdg602$  &  $0\fdg110$  &  71.1 (8.4$\sigma$)  &  38  &  $(15.1\pm3.7)\times10^{-7}$  &  $-2.08\pm0.19$  \\
57401.0--57402.2  &  $8\fdg969$  &  $61\fdg537$  &  $0\fdg063$  &  91.4 (9.6$\sigma$)  &  55  &  \phn$(5.5\pm1.4)\times10^{-7}$   &  $-1.86\pm0.15$  \\[+0.5ex]
\makecell[l]{57066.5--57066.9 and \\ 57401.0--57402.2} 
                  &  $8\fdg975$  &  $61\fdg548$  &  $0\fdg056$  &  154 (12.4$\sigma$)  &  94  &  \phn$(7.7\pm1.4)\times10^{-7}$   &  $-1.95\pm0.12$  \\
\enddata
\end{deluxetable*}


\section{X-Ray Data Analysis}

The region near Fermi J0035+6131 was observed with \textsl{XMM-Newton}
\citep{2001A&A...365L...1J} on 2016 January~24, 10~days after the
\textsl{Fermi} LAT detection.
We obtained \mbox{11.2 ks} of data from each of the EPIC MOS cameras \citep{2001A&A...365L..27T}
and \mbox{9.8 ks} from the EPIC PN camera \citep{2001A&A...365L..18S}.
All three cameras were operated in full frame mode with the medium blocking filters.
The \mbox{X-ray} data were filtered to include only good events
with patterns \mbox{0--12} for MOS and \mbox{0--4} for PN.
For our analysis we selected \mbox{X-ray} events in the energy range \mbox{0.5--12 keV}.

\begin{figure}
\includegraphics[width=\linewidth]{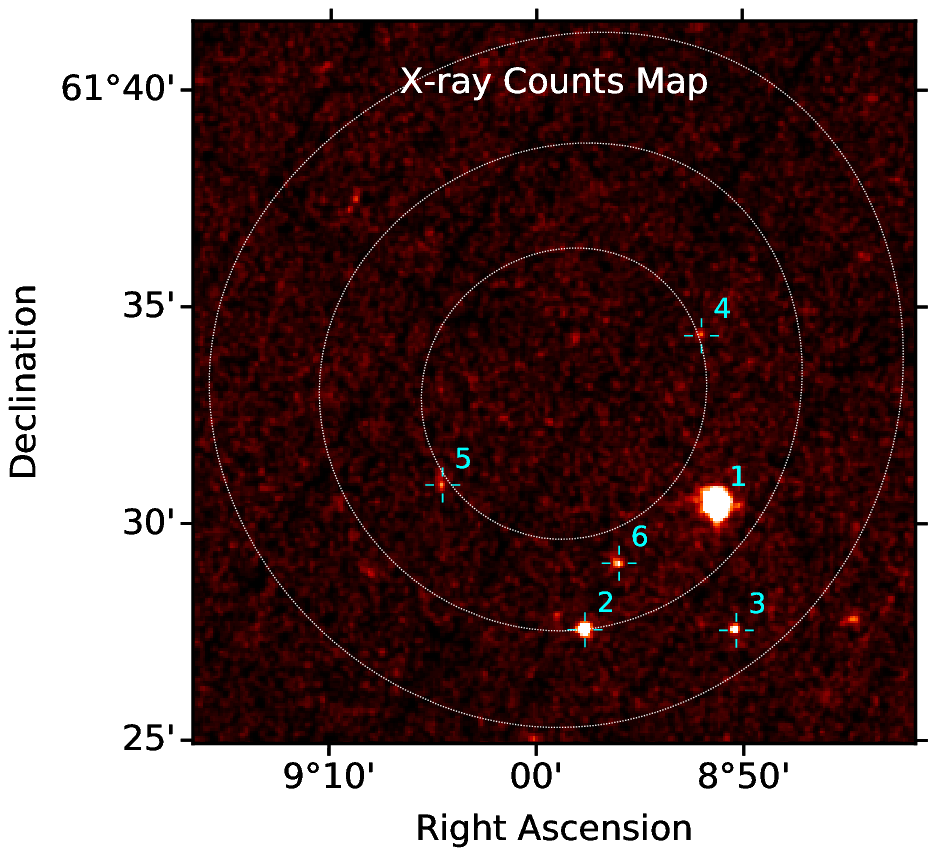}
\caption{
\mbox{X-ray} counts map (\mbox{0.5--12 keV}) of the region near Fermi J0035+6131
obtained with \textsl{XMM-Newton} 10 days after the first detection of the gamma-ray transient.
The contours represent the $1\sigma$, $2\sigma$, and $3\sigma$ (86\%, 95\%, and 99.7\%)
confidence regions of the gamma-ray source location obtained from the \textsl{Fermi} LAT data.
\mbox{X-ray} sources detected inside the $3\sigma$ confidence region are labeled with the IDs
shown in Table~\ref{xray-sources}).
\label{xray-image}}
\end{figure}

An \mbox{X-ray} counts map of the region near Fermi J0035+6131 is shown in
Figure~\ref{xray-image}.
The contours represent the confidence regions of the transient's location
we obtained from our analysis of the \textsl{Fermi} LAT data.
Six \mbox{X-ray} sources are detected with a likelihood DET\_ML$>$15 inside the $3\sigma$ (99.7\%)
\textsl{Fermi} LAT confidence region.
Source locations, \mbox{X-ray} fluxes, hardness ratios, and likely associations for these sources
are shown in Table~\ref{xray-sources}.

\begin{deluxetable*}{lcccccccccccl}
\tablecaption{
\mbox{X-ray} sources detected with \textsl{XMM-Newton} inside the $3\sigma$ (99.7\%) confidence
region of the gamma-ray source location of Fermi J0035+6131.
The source location has, in addition to the statistical error shown, a systematic error of $0\farcs5$.
\label{xray-sources}}
\tablehead{
ID  &  \colhead{}  &  \multicolumn{3}{c}{Source Location}  &  \colhead{}  &  \multicolumn{3}{c}{Flux ($\mathrm{10^{-14}\ erg\ cm^{-2}\ s^{-1}}$)}  &  \colhead{}  &  \colhead{Hardness Ratio}  &  \colhead{}  &  Likely Association (Offset)  \\[-1ex]
\cline{3-5} \cline{7-9} \\[-2.5ex]
  &  &  \colhead{R.A.}    &   \colhead{Decl.}  &  \colhead{Stat. Error}  &  &  \colhead{\makecell{$F_\mathrm{total}$ \\ (0.5--12 keV)}}  &  \colhead{\makecell{$F_\mathrm{soft}$ \\ (0.5--2 keV)}}  &  \colhead{\makecell{$F_\mathrm{hard}$ \\ (2--12 keV)}}  &  &  \colhead{$\frac{F_\mathrm{hard}-F_\mathrm{soft}}{F_\mathrm{total}}$}
}
\startdata
1  &  &  $8\fdg85510$  &  $61\fdg50846$  &  $0\farcs2$  &  &       98   &       13   &       85   &  &  \phs0.74\phs  &  &  VCS4 J0035+6130 ($0\farcs7$)  \\
2  &  &  $8\fdg96187$  &  $61\fdg45978$  &  $0\farcs5$  &  &      9.5   &      1.5   &      8.0   &  &  \phs0.70\phs  &  &  HD 3191 ($0\farcs1$)  \\
3  &  &  $8\fdg84006$  &  $61\fdg45949$  &  $0\farcs9$  &  &      3.3   &      0.3   &      3.0   &  &  \phs0.80\phs  &  &  \ldots  \\
4  &  &  $8\fdg86784$  &  $61\fdg57279$  &  $1\farcs5$  &  &  \phn1.14  &  \phn0.16  &  \phn0.98  &  &  \phs0.71\phs  &  &  \ldots  \\
5  &  &  $9\fdg07663$  &  $61\fdg51540$  &  $1\farcs2$  &  &  \phn0.70  &  \phn0.68  &  \phn0.02  &  &   $-0.96$\phs  &  &  USNO-B1  1515-0021314 ($0\farcs9$)  \\
6  &  &  $8\fdg93439$  &  $61\fdg48537$  &  $1\farcs3$  &  &  \phn0.63  &  \phn0.60  &  \phn0.03  &  &   $-0.91$\phs  &  &  USNO-B1  1514-0021151 ($1\farcs6$)  \\
\enddata
\end{deluxetable*}

Four of the detected \mbox{X-ray} sources exhibit a large hardness ratio suggesting an accreting object
or other type of high-energy source.
Their hard \mbox{X-ray} spectra make them potential candidates for the counterpart
of the gamma-ray source.
The two brightest of these sources are associated with the unidentified radio source
VCS4 J0035+6130 and the star \mbox{HD 3191}, respectively.
No counterparts at other wavelengths are known for the other two hard \mbox{X-ray} sources.
The two remaining sources exhibit a very soft \mbox{X-ray} spectrum, and they are unlikely
to be associated with the gamma-ray flare.
Both have a known optical counterpart and are likely \mbox{X-ray} active stars.

For the two brightest sources, we fitted the \mbox{X-ray} spectra with various models
using XSPEC \citep{1996ASPC..101...17A} and estimated the spectral parameter using
a maximum likelihood method.
In both cases, the spectra are equally well fit by either an absorbed power-law model
or an absorbed bremsstrahlung model.
However, an absorbed blackbody model does not provide a good fit and can be ruled out.
The best-fit parameters for the power-law and bremsstrahlung models are shown in
Table~\ref{xray-spec-table}.
For the other sources, the \mbox{X-ray} flux was too low to determine meaningful parameters
from a spectral fit.

\begin{deluxetable}{lcc}
\setlength{\tabcolsep}{0.7em}
\tablecaption{
\mbox{X-ray} spectral parameters for the two brightest \mbox{X-ray} sources,
VCS4 J0035+6130 and \mbox{HD 3191}, obtained by fitting two different spectral models,
an absorbed power-law model and an absorbed single-temperature bremsstrahlung model,
to the \textsl{XMM-Newton} spectra.
The flux shown is the absorbed flux in the \mbox{0.5--12 keV} energy range,
$N_\mathrm{H}$ is the neutral hydrogen column density \citep[using abundances by][]{2000ApJ...542..914W},
and uncertainties are shown at 90\% confidence.
Parameters were estimated using a maximum likelihood method with $C$~statistic
\citep{1979ApJ...228..939C}, and the $C$~value of the fits as well as
the number of degrees of freedom (dof) are shown.
\label{xray-spec-table}}
\tablehead{
\colhead{}  &  \colhead{VCS4 J0035+6130}  &  \colhead{HD 3191}
}
\startdata
\multicolumn{3}{c}{\textsl{Power-law Model}}  \\
Flux ($\mathrm{10^{-14}\ erg\ cm^{-2}\ s^{-1}}$)  &  $103\pm7$\phn\phn  &  $10.6\pm2.5$\phn  \\
Photon index                                 &  $-1.57\pm0.10$\phs  &  $-1.5\pm0.4$\phs  \\
$N_\mathrm{H}$ ($\mathrm{10^{21}\ cm^{-2}}$)               &  $7.5\pm1.0$     &  $4.5\pm3.0$  \\
$C$ value (dof)                            &  134 (120)       &  14.1 (24) \\
\hline
\multicolumn{3}{c}{\textsl{Bremsstrahlung Model}}  \\
Flux ($\mathrm{10^{-14}\ erg\ cm^{-2}\ s^{-1}}$)  &  $98\pm7$\phn    &  $10.1\pm2.7$\phn  \\
Temperature (keV)                            &  $15\pm5$\phn    &  $>6$  \\
$N_\mathrm{H}$ ($\mathrm{10^{21}\ cm^{-2}}$)               &  $6.6\pm0.8$     &  $4.1\pm2.2$  \\
$C$ value (dof)                            &  130 (120)       &  13.9 (24) \\
\enddata
\end{deluxetable}

\begin{figure}
\includegraphics[width=\linewidth]{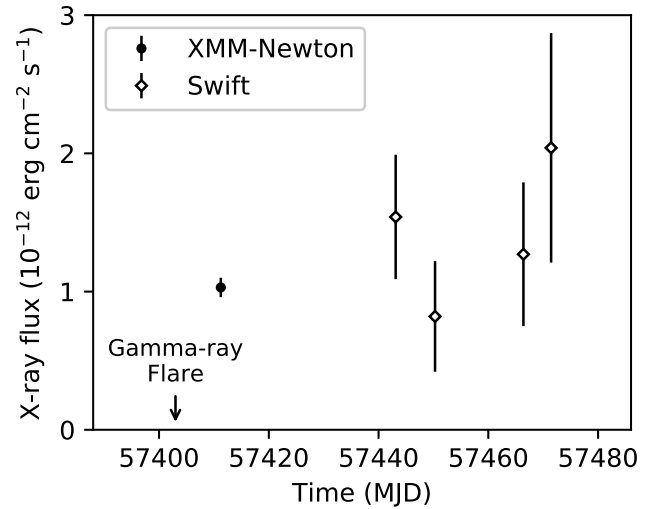}
\caption{
\mbox{X-ray} light curve of VCS4 J0035+6130 showing the flux in the \mbox{0.5--12 keV}
energy range during the \textsl{XMM-Newton} and \textsl{Swift} observations.
\label{xray-lc}}
\end{figure}

The region near Fermi J0035+6131 was also observed with the \textsl{Swift} XRT
\citep{2005SSRv..120..165B} four times between 42 and 70~days after
the \textsl{Fermi} LAT detection with individual exposures ranging
from 0.7 to \mbox{2.1 ks} in duration.
Because of the lower sensitivity of the XRT compared to \textsl{XMM-Newton}, we only detect
the brightest \mbox{X-ray} source, VCS4 J0035+6130.
We estimated the \mbox{X-ray} flux from this source for each \textsl{Swift} observation
by fitting the spectrum with an absorbed power-law model with the photon index
and neutral hydrogen column density fixed at the best-fit values obtained
from the \textsl{XMM-Newton} spectrum (Table~\ref{xray-spec-table}).
A more detailed spectral analysis was not possible because of the low number of photons
detected with the \textsl{Swift} XRT.
A light curve of our flux measurements for VCS4 J0035+6130 is shown in Figure~\ref{xray-lc}.
The \mbox{X-ray} brightness does not appear to be strongly variable or declining
as might be expected after a gamma-ray flare.
However, because of their large uncertainty, the \textsl{Swift} XRT measurements only provide poor
constraints on any potential \mbox{X-ray} variability.
For \mbox{HD 3191}, which was not detected with \textsl{Swift}, we determined an upper limit of
$26\times10^{-14}\ \mathrm{erg\ cm^{-2}\ s^{-1}}$ on the \mbox{0.5--12 keV} flux (\mbox{95\% c.l.}),
which is well above the flux determined from the \textsl{XMM-Newton} data.
Therefore, no conclusions can be drawn about potential \mbox{X-ray} variability of \mbox{HD 3191}.

\section{Discussion}

The high-energy processes occurring in gamma-ray emitting objects often lead to 
\mbox{X-ray} emission with a hard spectral slope.
The four \mbox{X-ray} sources with a large hardness ratio we detect inside
the \textsl{Fermi} LAT confidence region (Table~\ref{xray-sources})
could therefore be considered potential counterparts of Fermi J0035+6131.
However, for two of these sources, no counterparts are known at other wavelengths,
making it impossible to classify them without further multi-wavelength observations.
We therefore limit our discussion to the two brightest \mbox{X-ray} sources,
which have previously been detected at other wavelengths.

\subsection{VCS4 J0035+6130}

The brightest \mbox{X-ray} source we detect inside the \textsl{Fermi} LAT confidence region
is positionally coincident with the radio source VCS4 J0035+6130 \citep{2006AJ....131.1872P}
whose offset of $0\farcs7$ is comparable to the error of the \mbox{X-ray} source location
($0\farcs2$ statistical, $0\farcs5$ systematic).
VCS4 J0035+6130 (NVSS J003524+613030) has been observed in various radio surveys.
\citet{1998AJ....115.1693C} measured a radio flux density of \mbox{292 mJy} at \mbox{1.4 GHz},
\citet{1990A&AS...85..805F} a flux density of \mbox{230 mJy} at \mbox{2.7 GHz}, and
\citet{1991ApJS...75.1011G} a flux density of \mbox{180 mJy} at \mbox{4.85 GHz}.
\citet{atel8718} performed radio observations of the object \mbox{6--23 days} after
the gamma-ray flare and measured mean flux densities of 230, 210, and \mbox{350 mJy},
respectively, at 4.8, 11.2, and \mbox{21.7 GHz}.
These flux densities are higher than during past observations by a factor of \mbox{1.5--2}
at frequencies above \mbox{4.8 GHz}, which indicates significant variability.
The combined flux measurements suggest a spectral index of about $-0.5$.

\citet{2012A&A...544A..34P} analyzed VLBI data of VCS4 J0035+6130 and found that 80\%
of the observed radio flux is concentrated in a compact core with a size \mbox{$<$2 mas} (FWHM).
A total flux density of \mbox{110 mJy} was measured in the VLBI maps at both 2.3
and \mbox{8.6 GHz}, with \mbox{90 mJy} contained in the compact component.
This is somewhat lower than the flux densities obtained from the single-dish and shorter-baseline
observations mentioned above, which may indicate the presence of radio emission at angular scales
larger than the VLBI field of view.
The radio properties of VCS4 J0035+6130 are similar to those of the active galactic nuclei
investigated by \citet{2012A&A...544A..34P}, which generally exhibit milliarcsecond sizes
and spectral indices above $-0.5$.
The strong, compact radio emission, the flat radio spectrum, and the apparent radio flux
variability suggest that VCS4 J0035+6130 is an active galaxy serendipitously located
behind the Galactic disk.
The extragalactic nature of the object is supported by our estimates of the 
neutral hydrogen column density of $N_\mathrm{H}=(7.5\pm1.0)\times10^{21}\ \mathrm{cm^{-2}}$
and $N_\mathrm{H}=(6.6\pm0.8)\times10^{21}\ \mathrm{cm^{-2}}$ (depending on the spectral model),
which are somewhat larger than the total Galactic $N_\mathrm{H}$ along the line of sight
of $5.4\times10^{21}\ \mathrm{cm^{-2}}$ \citep{2005A&A...440..775K}.
While the larger $N_\mathrm{H}$ may indicate intrinsic absorption in the source,
the accuracy of our $N_\mathrm{H}$ measurements is insufficient to draw definitive conclusions.

In follow-up observations performed 14~days after the gamma-ray flare,
\citet{atel8706} discovered a near-infrared counterpart of VCS4 J0035+6130
with magnitudes $J=16.81$, $H=15.64$, and $Ks=14.86$.
The object had previously not been detected in the 2MASS survey \citep{2006AJ....131.1163S},
but \citet{atel8789} point out that this does not necessarily imply an increase
of the NIR flux.
No counterpart at visible wavelength was detected in the second Palomar Observatory
Sky Survey \citep{1991PASP..103..661R}, which has limiting magnitudes of
$B=22.5$, $R=20.8$, and $I=19.5$.
This non-detection may not be surprising given the strong extinction of visible light
from extragalactic objects near the Galactic plane.
Based on the dust maps by \citet{1998ApJ...500..525S}, we estimate that the total Galactic
extinction in the $V$~band along the line of sight is $A_V=3.9\ \mathrm{mag}$.
This implies limits on the intrinsic brightness of the object of
$B>17.4$, $R>17.6$, and $I>17.2$.
The NIR emission experiences significantly less extinction,
and the measurements by \citet{atel8706} correspond to intrinsic magnitudes of
$J=15.7$, $H=15.0$, and $Ks=14.4$.
Although the limits on the visual magnitudes are somewhat higher than the measured
NIR magnitudes, it cannot be concluded that the object experienced an increase
in brightness after the gamma-ray flare.
For example, observations by \citet{2011PASJ...63..639I} show larger visual
than NIR magnitudes with differences of $V-J=1.3\mathrm{-}2.8$
for blazars not significantly affected by extinction.
Therefore, the non-detection of VCS4 J0035+6130 at visible wavelengths
can be attributed to the strong extinction near the Galactic plane
without presuming variability of the source.

In summary, the strong, compact radio emission with a high spectral index,
the large neutral hydrogen column density,
and the non-detection at visible wavelengths strongly suggest that VCS4 J0035+6130
is an active galaxy serendipitously located behind the Galactic disk.
Other transient \textsl{Fermi} LAT sources at low Galactic latitudes have been identified
as extragalactic, such as Fermi J0109+6134 \citep{2010ApJ...718L.166V}.
\citet{2013ApJ...771...57A} found that most of the flaring gamma-ray sources
detected with \textsl{Fermi} near the Galactic plane are probably blazars.
VCS4 J0035+6130 is therefore the most likely counterpart of Fermi J0035+6131.
However, further optical observations, in particular redshift measurements, are needed
to confirm its extragalactic nature.

\subsection{HD 3191}

The second brightest \mbox{X-ray} source we detect inside the \textsl{Fermi} LAT
confidence region is positionally coincident with \mbox{HD 3191}, which is classified
as a \mbox{B1 IV:nn} star, i.e.\ a sub-giant with very broad absorption features.
The star has a visual magnitude of $V=8.6$ and an estimated distance of \mbox{$\sim$2 kpc}
based on a parallax measurement of $0.47\pm0.25\ \mathrm{mas}$ \citep{2016A&A...595A...2G}.
Our analysis of the \textit{XMM-Newton} data showed that \mbox{HD 3191} has an unusually hard
\mbox{X-ray} spectrum.
While early \mbox{B-type} stars are known to emit \mbox{X-rays} resulting from shocks
in theirs stellar winds, they generally exhibit much softer \mbox{X-rays} spectra
\citep[e.g.][]{2011MNRAS.416.1456O}.
From our spectral fit with a bremsstrahlung model, we derived a lower limit
of \mbox{6 keV} (\mbox{70 MK}) on the plasma temperature, which is significantly higher
than in typical early \mbox{B-type} stars.
However, the hard spectrum can be readily explained if \mbox{HD 3191} is in fact
an \mbox{X-ray} binary with a compact companion.
This interpretation is supported by recent radial velocity measurements.
\citet{atel8789} determined a heliocentric radial velocity of $-46.0\pm0.5\ \mathrm{km\ s^{-1}}$
on 2016 March~7, which differs from historical measurements by \citet{1961PDAO...12....1P}
who found $-22\pm3\ \mathrm{km\ s^{-1}}$.
Furthermore, in the optical spectrum only one stellar component is visible,
which hints at a compact companion if \mbox{HD 3191} is indeed part of a binary.

Based on our spectral fits, we estimate that \mbox{HD 3191} has an unabsorbed, bolometric
\mbox{X-ray} flux of $1.5\times10^{-13}$ $\mathrm{erg\ cm^{-2}\ s^{-1}}$ which corresponds
to a luminosity of $7\times10^{31}\ \mathrm{erg\ s^{-1}}$ for a distance of \mbox{2 kpc}.
This is significantly lower than the $10^{34}\mathrm{-}10^{38}\ \mathrm{erg\ s^{-1}}$ typically
observed for High-mass \mbox{X-ray} Binaries (HMXBs) in which the stellar wind from an
\mbox{O-type} or \mbox{B-type} star is captured by a neutron star or black hole
\citep[e.g.][]{2005MNRAS.362..879S}.
However, the low end of the HMXB luminosity function is not well explored,
and it is conceivable that the low luminosity is due to a large orbital separation
resulting in a low rate of accretion.
Further observations, in particular radial velocity measurements,
are needed to establish whether \mbox{HD 3191} is indeed a binary with a compact companion.

\textsl{Fermi} has detected variable GeV emission from several \mbox{X-ray} binaries,
such as the microquasar Cygnus \mbox{X-3} \citep{2009Sci...326.1512F} and the High-mass
\mbox{X-ray} Binaries \mbox{LS I +61 303} and \mbox{LS 5039} \citep{2009ApJS..183...46A}.
The peak gamma-ray flux observed for these binaries is comparable to the one we found
for Fermi J0035+6131 \citep[e.g.][Table~2]{2013A&ARv..21...64D}.
However, their \mbox{X-ray} emission is generally \mbox{1--2} orders of magnitude stronger
than that detected from \mbox{HD 3191}.
E.g., \citet{2011ApJ...736L..11A} found flaring in the long-period gamma-ray
binary \mbox{PSR B1259-63} with a peak gamma-ray flux of
$\sim$2$\times10^{-6}\ \mathrm{photons\ cm^{-2}\ s^{-1}}$ (\mbox{$>$100 MeV}),
while the average \mbox{X-ray} flux was $\sim$10$^{-11}\ \mathrm{erg\ cm^{-2}\ s^{-1}}$,
two orders of magnitude higher than the $1\times10^{-13}\ \mathrm{erg\ cm^{-2}\ s^{-1}}$
we found for \mbox{HD 3191}.
Furthermore, the gamma-ray emission from these binaries exhibits flaring on longer
time scales than Fermi J0035+6131 (tens of days) or shows a strong orbital modulation.
The gamma-ray binaries are also detected in the radio band,
whereas \mbox{HD 3191} does not have a known radio counterpart.
It therefore seems unlikely that \mbox{HD 3191} is the source of the observed
gamma rays even if it can be confirmed to be an \mbox{X-ray} binary.


\section{Conclusions}

Considering the abundance of flaring, extragalactic gamma-ray sources detected with
\textsl{Fermi} and considering the multi-wavelength properties of the object,
VCS4 J0035+6130 is the most likely counterpart of Fermi J0035+6131
and probably an active galaxy serendipitously located behind the Galactic disk.
While the potential \mbox{X-ray} binary \mbox{HD 3191} cannot be completely ruled out
as an alternative candidate, its properties compared to known gamma-ray binaries
make it unlikely to be the source of the observed gamma rays.
Further observations, in particular redshift measurements for VCS4 J0035+6130
and radial velocity measurements for \mbox{HD 3191}, are needed to conclusively establish
the nature of these objects.


\acknowledgments

D.P. acknowledges support from NASA Guest Investigator Grant 15AJ88G.
This work is based on observations obtained with \textit{XMM-Newton}, an ESA science mission
with instruments and contributions directly funded by ESA member states and NASA.

\facilities{\textsl{Fermi} (LAT), \textsl{XMM}, \textsl{Swift} (XRT)}

\software{Fermi Science Tools, XMM-Newton SAS, XSPEC}



\bibliographystyle{aasjournal}
\bibliography{fermij0035}

\begin{thebibliography}{}
\expandafter\ifx\csname natexlab\endcsname\relax\def\natexlab#1{#1}\fi
\providecommand{\url}[1]{\href{#1}{#1}}
\providecommand{\dodoi}[1]{doi:~\href{http://doi.org/#1}{\nolinkurl{#1}}}
\providecommand{\doeprint}[1]{\href{http://ascl.net/#1}{\nolinkurl{http://ascl.net/#1}}}
\providecommand{\doarXiv}[1]{\href{https://arxiv.org/abs/#1}{\nolinkurl{https://arxiv.org/abs/#1}}}

\bibitem[{{Abdo} {et~al.}(2009{\natexlab{a}}){Abdo}, {Ackermann}, {Ajello},
  {et~al.}}]{2009Sci...326.1512F}
{Abdo}, A.~A., {Ackermann}, M., {Ajello}, M., {et~al.} 2009{\natexlab{a}},
  Science, 326, 1512, \dodoi{10.1126/science.1182174}

\bibitem[{{Abdo} {et~al.}(2009{\natexlab{b}}){Abdo}, {Ackermann}, {Ajello},
  {et~al.}}]{2009ApJS..183...46A}
---. 2009{\natexlab{b}}, \apjs, 183, 46, \dodoi{10.1088/0067-0049/183/1/46}

\bibitem[{{Abdo} {et~al.}(2010){Abdo}, {Ackermann}, {Ajello},
  {et~al.}}]{2010Sci...329..817A}
---. 2010, Science, 329, 817, \dodoi{10.1126/science.1192537}

\bibitem[{{Abdo} {et~al.}(2011){Abdo}, {Ackermann}, {Ajello},
  {et~al.}}]{2011ApJ...736L..11A}
---. 2011, \apjl, 736, L11, \dodoi{10.1088/2041-8205/736/1/L11}

\bibitem[{{Acero} {et~al.}(2015){Acero}, {Ackermann}, {Ajello},
  {et~al.}}]{2015ApJS..218...23A}
{Acero}, F., {Ackermann}, M., {Ajello}, M., {et~al.} 2015, \apjs, 218, 23,
  \dodoi{10.1088/0067-0049/218/2/23}

\bibitem[{{Ackermann} {et~al.}(2013){Ackermann}, {Ajello}, {Albert},
  {et~al.}}]{2013ApJ...771...57A}
{Ackermann}, M., {Ajello}, M., {Albert}, A., {et~al.} 2013, \apj, 771, 57,
  \dodoi{10.1088/0004-637X/771/1/57}

\bibitem[{{Arnaud}(1996)}]{1996ASPC..101...17A}
{Arnaud}, K.~A. 1996, in ASP Conf.\ Ser., Vol. 101, Astronomical Data Analysis
  Software and Systems V, ed. {G.~H. Jacoby \& J. Barnes (San Francisco, CA:
  ASP)}, 17

\bibitem[{{Atwood} {et~al.}(2013){Atwood}, {Albert}, {Baldini},
  {et~al.}}]{2013arXiv1303.3514A}
{Atwood}, W., {Albert}, A., {Baldini}, L., {et~al.} 2013, ArXiv e-prints.
\newblock \doarXiv{1303.3514}

\bibitem[{{Atwood} {et~al.}(2009){Atwood}, {Abdo}, {Ackermann},
  {et~al.}}]{2009ApJ...697.1071A}
{Atwood}, W.~B., {Abdo}, A.~A., {Ackermann}, M., {et~al.} 2009, \apj, 697,
  1071, \dodoi{10.1088/0004-637X/697/2/1071}

\bibitem[{{Brown} {et~al.}(2016){Brown}, {Vallenari}, {Prusti},
  {et~al.}}]{2016A&A...595A...2G}
{Brown}, A.~G.~A., {Vallenari}, A., {Prusti}, T., {et~al.} 2016, \aap, 595, A2,
  \dodoi{10.1051/0004-6361/201629512}

\bibitem[{{Burrows} {et~al.}(2005){Burrows}, {Hill}, {Nousek},
  {et~al.}}]{2005SSRv..120..165B}
{Burrows}, D.~N., {Hill}, J.~E., {Nousek}, J.~A., {et~al.} 2005, \ssr, 120,
  165, \dodoi{10.1007/s11214-005-5097-2}

\bibitem[{{Carrasco} {et~al.}(2016){Carrasco}, {Recillas}, {Porras},
  {Chavushyan}, \& {Carraminana}}]{atel8706}
{Carrasco}, L., {Recillas}, E., {Porras}, A., {Chavushyan}, V., \&
  {Carraminana}, A. 2016, ATel, 8706

\bibitem[{{Cash}(1979)}]{1979ApJ...228..939C}
{Cash}, W. 1979, \apj, 228, 939, \dodoi{10.1086/156922}

\bibitem[{{Cheung} {et~al.}(2010){Cheung}, {Donato}, {Wallace}, {Corbet},
  {Dubus}, {Sokolovsky}, \& {Takahashi}}]{atel2487}
{Cheung}, C.~C., {Donato}, D., {Wallace}, E., {et~al.} 2010, ATel, 2487

\bibitem[{{Condon} {et~al.}(1998){Condon}, {Cotton}, {Greisen}, {Yin},
  {Perley}, {Taylor}, \& {Broderick}}]{1998AJ....115.1693C}
{Condon}, J.~J., {Cotton}, W.~D., {Greisen}, E.~W., {et~al.} 1998, \aj, 115,
  1693, \dodoi{10.1086/300337}

\bibitem[{{Dubus}(2013)}]{2013A&ARv..21...64D}
{Dubus}, G. 2013, \aapr, 21, 64, \dodoi{10.1007/s00159-013-0064-5}

\bibitem[{{Furst} {et~al.}(1990){Furst}, {Reich}, {Reich}, \&
  {Reif}}]{1990A&AS...85..805F}
{Furst}, E., {Reich}, W., {Reich}, P., \& {Reif}, K. 1990, \aaps, 85, 805

\bibitem[{{Gregory} \& {Condon}(1991)}]{1991ApJS...75.1011G}
{Gregory}, P.~C., \& {Condon}, J.~J. 1991, \apjs, 75, 1011,
  \dodoi{10.1086/191559}

\bibitem[{{Ikejiri} {et~al.}(2011){Ikejiri}, {Uemura}, {Sasada},
  {et~al.}}]{2011PASJ...63..639I}
{Ikejiri}, Y., {Uemura}, M., {Sasada}, M., {et~al.} 2011, \pasj, 63, 639,
  \dodoi{10.1093/pasj/63.3.327}

\bibitem[{{Jansen} {et~al.}(2001){Jansen}, {Lumb}, {Altieri},
  {et~al.}}]{2001A&A...365L...1J}
{Jansen}, F., {Lumb}, D., {Altieri}, B., {et~al.} 2001, \aap, 365, L1,
  \dodoi{10.1051/0004-6361:20000036}

\bibitem[{{Kalberla} {et~al.}(2005){Kalberla}, {Burton}, {Hartmann}, {Arnal},
  {Bajaja}, {Morras}, \& {P{\"o}ppel}}]{2005A&A...440..775K}
{Kalberla}, P.~M.~W., {Burton}, W.~B., {Hartmann}, D., {et~al.} 2005, \aap,
  440, 775, \dodoi{10.1051/0004-6361:20041864}

\bibitem[{{Mattox} {et~al.}(1996){Mattox}, {Bertsch}, {Chiang},
  {et~al.}}]{1996ApJ...461..396M}
{Mattox}, J.~R., {Bertsch}, D.~L., {Chiang}, J., {et~al.} 1996, \apj, 461, 396,
  \dodoi{10.1086/177068}

\bibitem[{{Munari} \& {Valisa}(2016)}]{atel8789}
{Munari}, U., \& {Valisa}, P. 2016, ATel, 8789

\bibitem[{{Oskinova} {et~al.}(2011){Oskinova}, {Todt}, {Ignace}, {Brown},
  {Cassinelli}, \& {Hamann}}]{2011MNRAS.416.1456O}
{Oskinova}, L.~M., {Todt}, H., {Ignace}, R., {et~al.} 2011, \mnras, 416, 1456,
  \dodoi{10.1111/j.1365-2966.2011.19143.x}

\bibitem[{{Pandel}(2017)}]{2017AIPC.1792d0037P}
{Pandel}, D. 2017, in AIP Conf.\ Ser., Vol. 1792, 6th International Symposium
  on High Energy Gamma-Ray Astronomy, ed. {F.~A. Aharonian, W. Hofmann, \&
  F.~M. Rieger (Melville, NY: AIP)}, 040037

\bibitem[{{Pandel} \& {Kaaret}(2016)}]{atel8783}
{Pandel}, D., \& {Kaaret}, P. 2016, ATel, 8783

\bibitem[{{Petrie} \& {Pearce}(1961)}]{1961PDAO...12....1P}
{Petrie}, R.~M., \& {Pearce}, J.~A. 1961, Publications of the Dominion
  Astrophysical Observatory Victoria, 12, 1

\bibitem[{{Petrov} {et~al.}(2006){Petrov}, {Kovalev}, {Fomalont}, \&
  {Gordon}}]{2006AJ....131.1872P}
{Petrov}, L., {Kovalev}, Y.~Y., {Fomalont}, E.~B., \& {Gordon}, D. 2006, \aj,
  131, 1872, \dodoi{10.1086/499947}

\bibitem[{{Pivato} {et~al.}(2016){Pivato}, {Buson}, \& {Razzan}}]{atel8554}
{Pivato}, G., {Buson}, S., \& {Razzan}, M. 2016, ATel, 8554

\bibitem[{{Pushkarev} \& {Kovalev}(2012)}]{2012A&A...544A..34P}
{Pushkarev}, A.~B., \& {Kovalev}, Y.~Y. 2012, \aap, 544, A34,
  \dodoi{10.1051/0004-6361/201219352}

\bibitem[{{Reid} {et~al.}(1991){Reid}, {Brewer}, {Brucato},
  {et~al.}}]{1991PASP..103..661R}
{Reid}, I.~N., {Brewer}, C., {Brucato}, R.~J., {et~al.} 1991, \pasp, 103, 661,
  \dodoi{10.1086/132866}

\bibitem[{{Scargle} {et~al.}(2013){Scargle}, {Norris}, {Jackson}, \&
  {Chiang}}]{2013ApJ...764..167S}
{Scargle}, J.~D., {Norris}, J.~P., {Jackson}, B., \& {Chiang}, J. 2013, \apj,
  764, 167, \dodoi{10.1088/0004-637X/764/2/167}

\bibitem[{{Schlegel} {et~al.}(1998){Schlegel}, {Finkbeiner}, \&
  {Davis}}]{1998ApJ...500..525S}
{Schlegel}, D.~J., {Finkbeiner}, D.~P., \& {Davis}, M. 1998, \apj, 500, 525,
  \dodoi{10.1086/305772}

\bibitem[{{Shtykovskiy} \& {Gilfanov}(2005)}]{2005MNRAS.362..879S}
{Shtykovskiy}, P., \& {Gilfanov}, M. 2005, \mnras, 362, 879,
  \dodoi{10.1111/j.1365-2966.2005.09320.x}

\bibitem[{{Skrutskie} {et~al.}(2006){Skrutskie}, {Cutri}, {Stiening},
  {et~al.}}]{2006AJ....131.1163S}
{Skrutskie}, M.~F., {Cutri}, R.~M., {Stiening}, R., {et~al.} 2006, \aj, 131,
  1163, \dodoi{10.1086/498708}

\bibitem[{{Str{\"u}der} {et~al.}(2001){Str{\"u}der}, {Briel}, {Dennerl},
  {et~al.}}]{2001A&A...365L..18S}
{Str{\"u}der}, L., {Briel}, U., {Dennerl}, K., {et~al.} 2001, \aap, 365, L18,
  \dodoi{10.1051/0004-6361:20000066}

\bibitem[{{Trushkin} {et~al.}(2016){Trushkin}, {Nizhelskij}, \&
  {Erkenov}}]{atel8718}
{Trushkin}, S.~A., {Nizhelskij}, N.~A., \& {Erkenov}, A. 2016, ATel, 8718

\bibitem[{{Turner} {et~al.}(2001){Turner}, {Abbey}, {Arnaud},
  {et~al.}}]{2001A&A...365L..27T}
{Turner}, M.~J.~L., {Abbey}, A., {Arnaud}, M., {et~al.} 2001, \aap, 365, L27,
  \dodoi{10.1051/0004-6361:20000087}

\bibitem[{{Vandenbroucke} {et~al.}(2010){Vandenbroucke}, {Buehler}, {Ajello},
  {et~al.}}]{2010ApJ...718L.166V}
{Vandenbroucke}, J., {Buehler}, R., {Ajello}, M., {et~al.} 2010, \apjl, 718,
  L166, \dodoi{10.1088/2041-8205/718/2/L166}

\bibitem[{{Wilms} {et~al.}(2000){Wilms}, {Allen}, \&
  {McCray}}]{2000ApJ...542..914W}
{Wilms}, J., {Allen}, A., \& {McCray}, R. 2000, \apj, 542, 914,
  \dodoi{10.1086/317016}

\end{thebibliography}

\end{document}